\def\ca{\check{a}}
\def\cb{\check{b}}
\def\cc{\check{c}}
\def\cd{\check{d}}
\def\ta{\tilde{a}}
\def\tb{\tilde{b}}
\def\tc{\tilde{c}}
\def\td{\tilde{d}}
\def\tA{\tilde{A}}
\def\tD{\tilde{D}}
\def\tGamma{\tilde{\Gamma}}
\def\ha{\hat{a}}
\def\hb{\hat{b}}
\def\hc{\hat{c}}
\def\hd{\hat{d}}
\def\bh{{\bf h}}
\def\bs{{\bf s}}
\def\bt{{\bf t}}
\def\bm{{\bf m}}
\def\bn{{\bf n}}
\def\oa{\bar{a}}
\def\od{\bar{d}}

\def\calG{{\cal G}}

\def\caS{{\cal S}}
\def\Bii{{\mbox{{\sf Z}\kern-0.38em{\sf Z}}}}
\def\Bnn{{\mbox{{\sf N}\kern-0.38em{\sf N}}}}
\def\Bcc{{\mbox{{\sf C}\kern-0.38em{\sf C}}}}
\def\DEF{:=}
\def\diag{{\mbox{\rm diag}}}
\def\deg{{\mbox{\rm deg}}}
\def\unity{{1\kern-0.35em1}}

\def\LieGL{{\mbox{\bf GL}}}
\def\det{{\mbox{\rm det}}}
\def\tr{{\mbox{\rm tr}}}

\def\Ad{{\mbox{\rm Ad}}}
\def\inv{^{-1}}
\def\GL{{\sf Gl}}
\def\GLM{\GL_{\rm -}}
\def\GLZ{\GL_{\rm 0}}
\def\GLP{\GL_{\rm +}}

\def\mod{\mathchoice{{\mbox{\rm mod}}}{{\mbox{\rm mod}}}{
{\mbox{\scriptsize\rm mod}}}{{\mbox{\scriptsize\rm mod}}} }
\def\GLRES{\GL_{\mbox{\scriptsize\rm res}}}
\def\GLKdV{\GL_{\mbox{\scriptsize\rm KdV}}}
\def\BEA{\begin{eqnarray}}
\def\EEA{\end{eqnarray}}
\def\BEQ{\begin{equation}}
\def\EEQ{\end{equation}}
\def\bref#1{(\ref{#1})}
\def\tmatrix#1#2#3#4{
    \left(\begin{array}{cc} #1 & #2 \\ #3 & #4 \end{array}\right)}
\def\Case#1#2#3#4{
    \left\{\begin{array}{l} #1,\, #2 \\ #3,\, #4 \end{array}\right.}
\documentstyle[12pt]{article}
\begin{document}
\title{Negative Flows of the KP-hierarchy}
\author{Guido Haak\thanks{supported by the Kansas Institute for Theoretical
and Computational Science at the University of Kansas}\\
Department of Mathematics\\ 405 Snow Hall\\
University of Kansas\\ Lawrence, KS 66045}
\date{}
\maketitle
\vskip-10cm
\noindent Report KITCS94-2-1 \\
hep-th/9403152
\vskip9cm
\begin{abstract}
We construct a Grassmannian-like formulation for the potential KP-hierarchy
including additional ``negative'' flows.
Our approach will generalize the notion of a
$\tau$-function to include negative flows.
We compare the resulting hierarchy with results by Hirota,
Satsuma and Bogoyavlenskii.
\end{abstract}
\section{Introduction}
It is a well known fact, that the modified KdV-hierarchy
allows an extension to a larger set of commuting
differential equations.
These are in fact the members
of the sine-Gordon hierarchy, as stated for example in \cite{DS}.
The additional commuting flows are still connected with the action
of an infinite dimensional abelian group on an infinite dimensional manifold
as it is used in the classical works of Adler and van~Moerbeke \cite{AvM} and
Segal and Wilson.
In the latter work, a part of a commutative group $\Gamma$ of
multiplication operators acts nontrivial on the quotient
$\GLRES/{\GLRES}_+$ of two infinite dimensional
groups by left multiplication.
The resulting quotient can be interpreted as an infinite dimensional
Grassmannian manifold. Out of this viewpoint Sato and later Segal and Wilson
derived exact analytic results for a certain class of solutions.

In this paper we investigate the flows of the potential KP-equation
in this setting, giving a formulation in terms of $\tau$-functions and
group actions on an infinite dimensional group. The extension of the
hierarchy is obtained by changing the investigated action of $\Gamma$ on
$\GLRES$ to conjugation, which extends the group of flows acting
nontrivial on a certain splitting subgroup of $\GLRES$.

Due to the natural $\Bii$-grading in the investigated hierarchy, we call the
additional flows ``negative'' flows.

The outline of this work follows the paper \cite{DNS89}, where the case of the
potential KP-equation is investigated:

In section~\ref{configsec} we take as configuration space of the extended
integrable system the whole group $\GLRES$ \cite{SW85},
which replaces the Grassmannian. This group is split into three subgroups
represented as lower triangular, diagonal and upper triangular block matrices.

Letting the abelian group $\Gamma$
of Segal and Wilson act by conjugation we obtain in section~\ref{flowsec}
a nontrivial action of the {\em whole} group $\Gamma$.

In section~\ref{matrixsec} we obtain differential equations relating the
matrix elements of the block matrices.

With a suitably defined $\tau$-function (section~\ref{tausec}) we
show in section~\ref{hiersec}, that out of these actions new scalar equations
result, extending the potential KP-hierarchy.

We also give some thought to the set of solutions, which in
section~\ref{solsec} is shown to
be much more complicated than in the potential KP case, which was investigated
in \cite{DPL1}.

In the last chapter we compute the $LU(2)$ reduction of the flows, yielding
an equation of Hirota and Satsuma, which commutes with the potential
KdV-equation. This equation was also investigated by Bogoyavlenskii.

\section{The configuration space} \label{configsec}
We start out by recalling the setting of Segal and Wilson \cite{SW85}.

Let $H$ be an infinite dimensional separable complex Hilbert space together
with a splitting
\BEQ
H=H_+\oplus H_- \label{split}
\EEQ
into the orthogonal direct sum of two infinite dimensional closed
subspaces $H_+$ and $H_-$.

We look at the group $\GLRES$ defined also by Segal and Wilson, of
invertible matrices which take w.r.t.\ the splitting \bref{split} the form
\BEQ
\left(\begin{array}{cc} a & r \\ s & d
\end{array}\right)
\EEQ
where $a$ and $d$ are Fredholm operators, and $r$ and $s$ are
compact. The connected components of $\GLRES$ are labelled by the
Fredholm index of the upper (or lower) diagonal block.

We now look at a factorization of elements of $\GLRES$
into diagonal, upper and lower triangular block matrices in the
following way: We split every element $g$ in $\GLRES$ as
\BEQ\label{splitting}
g=\left(\begin{array}{cc} A & B \\ C & D \end{array} \right)=
\left(\begin{array}{cc} \unity & 0 \\ c & \unity \end{array} \right)
\left(\begin{array}{cc} a & 0 \\ 0 & d \end{array} \right)
\left(\begin{array}{cc} \unity & b \\ 0 & \unity \end{array} \right)=
\left(\begin{array}{cc} a & ab \\ ca & cab+d \end{array} \right)
\EEQ
if this is possible.
Here the blocks $b$ and $c$ are again compact
and the diagonal blocks are invertible Fredholm.
Explicitely the splitting reads:
\BEA
a & = & A \nonumber\\
b & = & A\inv B \nonumber\\
c & = & CA\inv \nonumber\\
d & = & D-CA\inv B
\EEA
We therefore have a splitting into three subgroups of $\GLRES$, which we
denote by $\GLM$, $\GLZ$ and $\GLP$.

The set of elements of $\GLRES$ which are splittable is
characterized by the invertibility of the upper diagonal
block, thereby forming a dense open subset of $\GLRES$. This will lateron
motivate the definition of the
$\tau$-function in our setting without reference to a Grassmannian.

\section{The flows} \label{flowsec}
The representation of $\GLRES$ as a space of block matrices inherits
a choice of basis for the separable vector space $H$. We
number the basis vectors by integer numbers. Let $e_j$, $j\geq0$, and
$e_j$, $j<0$, be the basis vectors of $H_+$ and $H_-$, respectively.

We let $\GLRES$ act on itself by conjugation.
and choose the following generators of action:\\
Let $\Lambda$ be the double
right shift, i.e.\ the operator mapping $e_j$ to $e_{j+1}$. It is
represented by the matrix
\BEQ
\Lambda=\sum_{j\in\Bii}e_{j,j+1},
\EEQ
where $e_{j,k}$ are the matrix units
$(e_{j,k})_{l,m}=\delta_{j,l}\cdot\delta_{k,m}$.

We define the generator of the $m$-th flow to be $\Lambda^m$. Here $m$
can be any nozero integer.
Then the flows on $\GLRES$ are defined by
\BEQ \label{action}
g(\bt)=\exp(\sum_{m\neq0} t_m\Lambda^m)g(0)\exp(-\sum_{m\neq0}
t_m\Lambda^m).
\EEQ
$\bt$ denotes the vector with coordinates $t_j$, $j\in\Bii$ and $g(0)$ is an
initial value for the flow.

As long as only finitly many $t_j$ are nonzero, the flow on $\GLRES$
is obviously continuous, which results in the definition of continuous
local flows on the three factors $\GLM$, $\GLZ$ and $\GLP$ in
\bref{splitting}.
\BEQ \label{action2}
g_-(\bt)g_0(\bt)g_+(\bt)\DEF g(\bt).
\EEQ
It is easy to see, that none of these flows acts trivially on $\GLRES$ or
either of the splitting subgroups $\GLP$ and $\GLM$.

Our setting generalizes the Grassmannian flows:
Calculating the splitting
explicitly we will see, that the negative (positive) flows act
by simple translation on $\GLM$ ($\GLP$) and therefore negative flows are not
of interest on (the big cell of) the Grassmannian $\cong\GLRES/\GLZ\GLP$.

For later use we also define the abelian group $\Gamma\subset\GLRES$
generated by
the exponentials of the matrices $\Lambda^m$, $m\in\Bii$, and its subgroups
$\Gamma_+$ and $\Gamma_-$  generated by the exponentials of
the matrices $\Lambda^m$ for $m\geq 0$ and $m\leq0$, respectively. Obviously
$\Gamma=\Gamma_-\Gamma_+$. These groups of course coincide with the groups
defined in \cite{SW85}.

\section{Matrix equations} \label{matrixsec}
{}From now on we write for the matrix $\exp(\sum_m t_m\Lambda^m)$ and its
inverse
${a\,b\choose c\,d}$ and ${\ha\,\hb\choose\hc\,\hd}$,
respectively. In addition we write $\ca$, $\cb$, $\cc$, $\cd$
for the matrix elements of the
splitting \bref{splitting} at $\bt$ and $\ta$, $\tb$, $\tc$, $\td$
for the matrix elements of the splitting at $\bt+\bh$,
$\bh=(\ldots,h_{-1},h_1,\ldots)$.
Clearly for the beginning we consider only $g(\bt)$ which are splittable.
Then $g(\bt+\bh)$ is splittable for $\bh$ (i.e.\ all $h_i$)
sufficiently small. In fact, as
we will see later, $g(\bt)$ is not splittable only for isolated values of
$\bt$.

Equations \bref{action} and \bref{action2} lead to the following
block matrix equations:
\BEA
\lefteqn{\left(\begin{array}{cc} a & b \\ c & d \end{array}\right)
\left(\begin{array}{cc} \unity & 0 \\ \cc & \unity \end{array}\right)
\left(\begin{array}{cc} \ca & 0 \\ 0 & \cd \end{array}\right)
\left(\begin{array}{cc} \unity & \cb \\ 0 & \unity \end{array}\right)
\left(\begin{array}{cc} \ha & \hb \\ \hc & \hd \end{array}\right)
} & & \nonumber\\
& = & \kern-5pt\left(\begin{array}{cc}
a\ca\ha+b\cc\ca\ha+a\ca\cb\hc+b(\cc\ca\cb+\cd)\hc &
(a\ca\cb+b(\cc\ca\cb+\cd))\hd + (a+b\cc)\ca\hb \\
c\ca(\ha+\cb\hc) + d(\cc\ca\ha+(\cc\ca\cb+\cd)\hc) &
d(\cc\ca\cb+\cd)\hd + c\ca\cb\hd + d\cc\ca\hb + c\ca\hb \end{array}\right)
\nonumber\\
& = & \left(\begin{array}{cc} \ta & \ta\tb \\ \tc\ta & \td+\tc\ta\tb
\end{array}\right). \label{matrixflow}
\EEA
Notice that $\ha=a\inv$ and $\hd=d\inv$.

Regarding the matrix ${\ca \, \cb\choose \cc \, \cd}$ as a function of the
flow parameter $\bt$ we easily derive a set of differential equations for the
blocks. To write them in a compact notation, we define the following
abbreviation for the blocks of the matrix $\Lambda^m$, $m>0$:
\BEA
\Lambda^m & = &
\tmatrix{\Lambda^m_{++}}{\Lambda^m_{+-}}0{\Lambda^m_{--}} \\
\Lambda^{-m} & = & \tmatrix{\Lambda^{-m}_{++}}0{\Lambda^{-m}_{-+}}
{\Lambda^{-m}_{--}}.
\EEA
Furtheron we will also use the subscripts $++$, $+-$, $-+$ and $--$ to
denote the blocks of a matrix.

The blocks of $\Lambda^m$ are given by
\BEA
\Lambda^m_{++} & = & \Case{\sum_{k\geq0} e_{k+m,k}}{m>0}{\sum_{k\geq0}
e_{k,k-m}}{m<0}, \\
\Lambda^m_{+-} & = & \sum_{k=0}^{m-1} e_{k,k-m},\, m>0, \\
\Lambda^m_{--} & = & \Case{\sum_{k<0} e_{k,k-m}}{m>0}{\sum_{k<0}
e_{k+m,k}}{m<0}, \\
\Lambda^m_{-+} & = & \sum_{k=0}^{-m-1} e_{k+m,k},\, m<0.
\EEA
Denoting with $\partial_m$ the partial derivative w.r.t.\ the
parameter $t_m$, we end up with the equations:
\BEA
\partial_m\ca & = & \Lambda^m_{++}\ca-\ca\Lambda^m_{++}
+\Case{\Lambda^m_{+-}\cc\ca}{m>0}{-\ca\cb\Lambda^m_{-+}}{m<0}, \label{ap}\\
\partial_m(\ca\cb) & = & \Lambda^m_{++}\ca\cb-\ca\cb\Lambda^m_{--}
+\Case{\Lambda^m_{+-}(\cc\ca\cb+\cd)-\ca\Lambda^m_{+-}}{m>0}0{m<0},
\label{abeq}\\
\partial_m(\cc\ca) & = & \Lambda^m_{--}\cc\ca-\cc\ca\Lambda^m_{++}
+\Case{0}{m>0}{\Lambda^m_{-+}\ca-(\cc\ca\cb+\cd)\Lambda^m_{-+}}{m<0},
\label{caeq}\\
\partial_m\cd & = & \Lambda^m_{--}\cd-\cd\Lambda^m_{--}
+\Case{\cc(\ca\Lambda^m_{+-}-\Lambda^m_{+-}\cd)}{m>0}{(\cd\Lambda^m_{-+}-
\Lambda^m_{-+}\ca)\cb}{m<0}.
\EEA
{}From this we get the equations for the offdiagonal blocks $\cb$ and
$\cc$:
\BEA
\partial_m\cb & = & \Lambda^m_{++}\cb-\cb\Lambda^m_{--}
-\Case{\Lambda^m_{+-}-\ca\inv\Lambda^m_{+-}\cd}{m>0}{
-\cb\Lambda^m_{-+}\cb}{m<0}, \label{cbeq}\\
\partial_m\cc & = & \Lambda^m_{--}\cc-\cc\Lambda^m_{++}
-\Case{\cc\Lambda^m_{+-}\cc}{m>0}{
\cd\Lambda^m_{-+}\ca\inv-\Lambda^m_{-+}}{m<0}. \label{cceq}
\EEA
For $m>0$, \bref{cceq} is the Riccati type differential equation
already encountered in \cite{DNS89}, which has a negative counterpart,
\bref{cbeq}, for $m<0$.
In coordinates they read ($i<0$, $j\geq0$, $m>0$):
\BEA
\partial_m\cc_{i,j} & = &
\cc_{i-m,j}-\cc_{i,j+m}-\sum_{k=0}^{m-1}\cc_{i,k}\cc_{k-m,j},
\label{cceqcoord}\\
\partial_{-m}\cb_{j,i} & = &
\cb_{j+m,i}-\cb_{j,i-m}+\sum_{k=0}^{m-1}\cb_{j,k-m}\cb_{k,i}.
\label{cbeqcoord}
\EEA
Therefore each of the offdiagonal blocks is given by its first row {\em
or} column.

\section{The $\tau$-function} \label{tausec}
In order to motivate the definition of the $\tau$-function in our
setting, we consider for every $k\in\Bnn$ the subgroup $\GLRES(k)$ of $\GLRES$
of invertible block matrices
w.r.t.\ the splitting \bref{split}, which are of the form
\BEQ
\left(\begin{array}{cc} \unity+p & r \\ s & \unity+q
\end{array}\right)
\EEQ
where $p$ and $q$ are operators of trace class, $r$ is in the
$k$-th Schatten ideal in $B(H_-,H_+)$ and $s$ is in the ${k\over k-1}$-th
Schatten ideal in $B(H_+,H_-)$.
These groups are proper connected subgroups of $\GLRES$ for every
parameter $k$.

As the elements of $\GLRES(k)$ have the form
$$
\unity+\tmatrix{A}{B}{C}{D},
$$
where the blocks $A$, $B$, $C$, $D$ take values in different ideals of
operators, the action \bref{action} induces an action of $\GLRES(k)$ on itself
for every $k$.
The splittable elements of $\GLRES(k)$ are those for which $p$ has no
eigenvalue $-1$. Also in this case all factors of \bref{splitting} are in
$\GLRES(k)$.

The definition of $\GLRES(k)$ allows us to define the following
complex function on $\GLRES(k)$: For a fixed splittable element
$g_0$ in $\GLRES(k)$, the function
$\tau_{g_0}:\GLRES(k)\longrightarrow\Bcc$ is defined as the quotient
of the determinant of the upper diagonal blocks of $gg_0g\inv$ and
$g_0$, i.e.\
$$
\tau_{g_0}(g)\DEF{\det((gg_0g\inv)_{++})\over
\det(g_{++}(g_0)_{++}(g\inv)_{++})}.
$$
By the remark at the end of section~\ref{configsec},
this function is nonzero precisely when $g$ is splittable.

In terms of blocks the $\tau$-function is given by
\BEA
\tau_{g_0}(g) & = &
\det((a+b\cc)\ca(\ha+\cb\hc)+b\cd\hc)\det(a\ca\ha)\inv \nonumber\\
& = & \det(\unity+(a\inv b\cc\ca+\ca\cb\hc\ha\inv+a\inv
b(\cc\ca\cb+\cd)\hc\ha\inv)\ca\inv) \nonumber\\
& = & \det(\unity+\ca\inv(a\inv b\cc\ca+\ca\cb\hc\ha\inv+a\inv
b(\cc\ca\cb+\cd)\hc\ha\inv)), \label{taudef}
\EEA
which reduces to the known expression \cite{SW85,DNS89} if we look only
at elements $g$ describing positive flows and to a similar expression
for purely negative flows.

The offdiagonal blocks $b$ and $\hc$ of the action matrices and therefore all
product terms in the last two lines of \bref{taudef} are of trace class.
Thus the last two lines in \bref{taudef} are well defined and equal
for all splittable elements $g$ of $\GLRES$.
We use them to define the $\tau$-function on the whole of $\GLRES$. If $g$ is
not splittable it is natural to set $\tau_{g_0}(g)=0$.

If $g=g(\bt)=\exp(\sum_mt_m\Lambda^m)$ we also write
$$
\tau(\bt,g_0)\DEF\tau_{g_0}(g).
$$
We will see lateron, that this function is analytic as a function from $\Gamma$
to $\Bcc$\kern1mm.

{}From its definition $\tau$ inherits an equivariance property w.r.t.\
the used group action. If $\bt$ and $\bs$ are both negative or both
positive flows, then as usual
\BEQ
\tau(\bt+\bs,g_0) = \tau(\bs,g(\bt))\cdot\tau(\bt,g_0).
\EEQ
Otherwise we easily get by explicit calculation using \bref{taudef} and
\bref{matrixflow}:
\BEA
\tau(\bt_++\bs_-,g_0) &  = & \tau(\bt_+,g(\bs_-))\cdot\tau(\bs_-,g_0)\cdot
c(g_-,g_+) \\
& = & \tau(\bs_-,g(\bt_+))\cdot\tau(\bt_+,g_0)\cdot c(g_-,g_+).
\label{groupprop}
\EEA
Here
$$
g_+=g_+(\bt_+)=\exp(\sum_{m>0}(\bt_+)_m\Lambda^m)
$$
describes a positive,
$$
g_-=g_-(\bs_-)=\exp(\sum_{m<0}(\bs_-)_m\Lambda^m)
$$
a negative flow and
$c(g_-,g_+)$ is the ``projective multiplier'' introduced by Segal and
Wilson \cite[Prop.~3.6]{SW85}:
\BEQ
c(g_-,g_+) = \det(a_+ a_- a_+\inv a_-\inv),
\EEQ
$a_+$ and $a_-$ being the upper diagonal blocks of $g_+$ and $g_-$,
respectively.

Using the fact that $\Gamma_-$ and $\Gamma_+$ act in a noncommuting
way on the determinant line bundle over the Grassmannian, they showed
that $c(g_-,g_+)$ is a homomorphism in every argument.
Therefore
\BEQ
c(g_-=e^{f_-},g_+=e^{f_+})=e^{S(f_-,f_+)},
\EEQ
with
\BEQ
S(f_-,f_+)=-\sum_m a_m b_m,
\EEQ
if
$$
f_-=\sum_{m>0}a_m\Lambda^{-m}
$$
and
$$
f_+=\sum_{m>0}b_m\Lambda^m.
$$
If we complexify the flow variables, it follows from the definition and
the known theorems on the
$\tau$-function in the Grassmannian setting, that it is holomorphic as a
function from $\Gamma_+$ to $\Bcc$ and as a function of $\Gamma_-$ of
$\Bcc$\kern1mm.
Obviously, if we work in the setting of Segal and Wilson,
$\tau(\bt,g_0)$ is also locally bounded and thus holomorphic
as a map from $\Gamma$, being the product of the abelian groups $\Gamma_-$
and $\Gamma_+$, endowed with the operator norm on the diagonal and the trace
norm on the offdiagonal blocks.
The zeros of $\tau$ are therefore complex hypersurfaces of (real) codimension
$2$ in $\Bcc\kern1mm^2$.

The first row of the lower triangular block $\cc$ and the first column
of the upper triangular block $\cb$ are given by the $\tau$
function. This is clear for $\cc$, as all calculations reduce to those
in \cite{DNS89} for positive flows.
Introducing the special element
$n_\zeta=\unity-\Lambda/\zeta=\exp(\sum_{m>0}(\bn_\zeta)_m\Lambda^m)$,
$|\zeta|>1$, in $\Gamma_+$, $(\bn_\zeta)_k=-k\inv\zeta^{-k}$, $k>0$, we get
\BEQ
\tau(\bn_\zeta,g(\bt))=1-\sum_{k=1}^\infty\zeta^{-k}\cc_{-1,k-1}(\bt).
\EEQ
For negative flows everything
translates analogously. We choose the special element
$m_\xi=(1-\xi/\Lambda)\inv=\exp(\sum_{m<0}(\bm_\xi)_m\Lambda^m)$, $|\xi|<1$,
in $\Gamma_-$. Then, with
$(\bm_\xi)_k=k\inv\xi^k$, $k<0$,
\BEA
\tau(\bm_\xi,g(\bt)) & = & \det(\unity+\ha\inv(\bm_\xi)\cb(\bt)
\hc(\bm_\xi)) \nonumber\\
& = & \det(\unity-(\unity-\xi\Lambda\inv_{++})\inv\cb(\bt)\Lambda\inv_{-+}\xi)
\nonumber\\
& = & 1-\sum_{k=1}^\infty \xi^k\cb_{k-1,-1}(\bt).
\EEA
Let $\bt=\bt_++\bt_-$ be such that $(\bt_+)_k=0$ for $k\leq0$ and
$(\bt_-)_k=0$ for $k\geq0$. The following calculation is analogous to
the one in \cite[section~5.8]{DNS89} ($m>0$):
\BEA
\lefteqn{\partial_{-m}\tau(\bt,g_0)} & & \nonumber\\
& = & \partial_{t_{-m}}|_{t_{-m}=0}\tau(\bt+t_{-m},g_0) \nonumber\\
& = & \partial_{t_{-m}}|_{t_{-m}=0}(\tau(\bt_++(\bt_-+t_{-m}),g_0)) \nonumber\\
& = & \partial_{t_{-m}}|_{t_{-m}=0}(\tau(\bt_-+t_{-m},g(\bt_+))
c(g_-(\bt_-+t_{-m}),g_+(\bt_+))\tau(\bt_+,g_0)) \nonumber\\
& = & \partial_{t_{-m}}|_{t_{-m}=0}(\tau(t_{-m},g(\bt))\tau(\bt_-,g(\bt_+))
\tau(\bt_+,g_0)c(g_-(\bt_-+t_{-m}),g_+(\bt_+))) \nonumber\\
& = & \partial_{t_{-m}}|_{t_{-m}=0}(\tau(t_{-m},g(\bt))\tau(\bt,g_0)
c(g_-(t_{-m}),g_+(\bt_+))).
\EEA
It follows by the same reasoning as before,
\BEA
\lefteqn{\partial_{-m}\ln(\tau(\bt,g_0))} & & \nonumber\\
& = & \partial_{t_{-m}}|_{t_{-m}=0}(\det(\unity+\ha\inv(t_{-m})
\cb(\bt)\hc(t_{-m}))-c(g_-(t_{-m}),g_+(\bt_+))) \nonumber\\
& = &
\partial_{t_{-m}}|_{t_{-m}=0}(\tr(\cb(\bt)\hc(t_{-m}))-c(g_-(t_{-m}),g_+(\bt_+))).
\EEA
Explicitly this reads:
\BEA
\partial_{-m}\ln(\tau(\bt,g_0)) & = &
-\tr(\cb(\bt)\Lambda^{-m}_{-+})-mt_m \nonumber\\
& = & -\sum_{k=1}^m\cb_{m-k,-k}-mt_m,\, m>0,
\EEA
analogous to
\BEA
\partial_m\ln(\tau(\bt,g_0)) & = & \tr(\Lambda^m_{+-}\cc(\bt))-mt_{-m}
\nonumber\\
& = & \sum_{k=1}^m \cc_{-k,m-k}-mt_{-m},\, m>0.
\EEA
Especially, together with equation \cite[(5.8.1)--(5.8.2)]{DNS89} we
have
\BEQ
\partial_{-1}\cc_{-1,0}=\partial_{-1}\partial_1\ln\tau+1=-\partial_1\cb_{0,-1}.
\label{bcrel}
\EEQ
This connects the two offdiagonal blocks and at the same time proves,
that the functions $\cc_{-1,0}$ and $\cb_{0,-1}$ are
meromorphic w.r.t.\ the variables $t_1$ and $t_{-1}$, being quotients
of holomorphic functions.
We will see lateron how far $\tau$ is determined by $\cc_{-1,0}$ and
$\cb_{0,-1}$.

\section{Extension of the potential KP-hierarchy} \label{hiersec}
{}From \cite{SW85,DNS89} we expect $\cc_{-1,0}$ to be the function
satisfying the (potential) KP-equation.
We first look at the equations for $\cc(\bt+\bn_\zeta)$ and
$\cb(\bt+\bm_\xi)$.
\BEA
\cc(\bt+\bn_\zeta)(a(\bn_\zeta)+b(\bn_\zeta)\cc(\bt)) & = &
d(\bn_\zeta)\cc(\bt) \\
(\ha(\bm_\xi)+\cb(\bt)\hc(\bm_\xi))\cb(\bt+\bm_\xi) & = &
\cb(\bt)\hd(\bm_\xi).
\EEA
To get differential equations, we develop $\cb(\bt+\bn_\zeta)$ and
$\cc(\bt+\bm_\xi)$ w.r.t.\ the parameters $\zeta$ and $\xi$, respectively.
\BEA
\cc(\bt+\bn_\zeta) & = & \sum_{k=0}^\infty \zeta^{-k} P_k\cc(\bt),\\
\cb(\bt+\bm_\xi) & = & \sum_{k=0}^\infty \xi^k P_{-k}\cb(\bt).
\EEA
$P_k$, $k\in\Bii$ are differential operators determined by the special
structure of  the elements $n_\zeta$ and $m_\xi$.
For example
\BEA
\vdots & & \nonumber\\
P_{-3} & = &
{1\over3}\partial_{-3}+{1\over2}\partial_{-1}\partial_{-2}+{1\over
3!}\partial_{-1}^3, \\
P_{-2} & = & {1\over2}(\partial_{-1}^2+\partial_{-2}), \\
P_{-1} & = & \partial_{-1}, \\
P_0 & = & 1, \\
P_1 & = & -\partial_1, \\
P_2 & = & {1\over2}(\partial_1^2-\partial_2), \\
P_3 & = & -{1\over3}\partial_3+{1\over2}\partial_1\partial_2-{1\over
3!}\partial_1^3, \\
\vdots & & \nonumber
\EEA
and in general for $k>0$:
$$
P_{-k}(\partial_{-1},\ldots,\partial_{-k})=-P_k(\partial_1,\ldots,\partial_k).
$$
With $b(\bn_\zeta)=-\zeta\inv\Lambda_{+-}$,
$a(\bn_\zeta)=\unity-\zeta\inv\Lambda_{++}$,
$\hc(\bm_\xi)=-\xi\Lambda\inv_{-+}$ and
$\ha(\bm_\xi)=\unity-\xi\Lambda\inv_{++}$ we get
\BEA
\sum_{k\geq0}\zeta^{-k}P_k\cc(\unity-\zeta\inv\Lambda_{++}-
\zeta\inv\Lambda_{+-}\cc) & = & \cc-\zeta\inv\Lambda_{--}\cc,\\
(\unity-\xi\Lambda\inv_{++}-\xi\cb\Lambda\inv_{-+})\sum_{k\geq0}\xi^kP_{-k}\cb
& = & \cb-\xi\cb\Lambda\inv_{--},
\EEA
or in coordinates:
\BEA
P_{k+1}\cc_{-1,j}-P_k\cc_{-1,j+1}-(P_k\cc_{-1,0})\cc_{-1,j} & = & 0,
\label{cmhier}\\
P_{-k-1}\cb_{i,-1}-P_{-k}\cb_{i+1,-1}-\cb_{i,-1}P_{-k}\cb_{0,-1} & = & 0,
\label{bphier}
\EEA
for $k\geq1$, $j\geq0$.
Multiplying the first equation with $\zeta^{-j-1}$, the second with
$\xi^{i+1}$, we end up after summation with the following equations ($k\geq1$):
\BEA
(P_{k+1}-\zeta P_k-(P_k\cc_{-1,0}))\tau(\bn_\zeta,g(\bt)) & = & 0, \\
(P_{-k-1}-\xi\inv P_{-k}-(P_{-k}\cb_{0,-1}))\tau(\bm_\xi,g(\bt)) & = & 0.
\EEA
Using the equations
\BEA
\ca(\bt+\bm_\zeta) & = & a(\bm_\zeta)\ca(\bt)(\ha(\bm_\zeta)+
\cb(\bt)\hc(\bm_\zeta)),\\
\ca(\bt+\bn_\xi) & = & (a(\bn_\xi)+b(\bn_\xi)\cc(\bt))\ca(\bt)\ha(\bn_\xi),
\EEA
which follow from \bref{matrixflow}, we get as a byproduct in the same way
explicit expressions for $P_k\ca$ ($k\geq1$):
\BEA
P_k\ca & = & (\ca\Lambda_{++}-\Lambda_{++}\ca-\Lambda_{+-}\cc\ca)
\Lambda_{++}^{k-1}\nonumber\\
& = & -\partial_1\ca\Lambda_{++}^{k-1},
\EEA
and
\BEA
P_{-k}\ca & = & \Lambda^{1-k}_{++}(\Lambda\inv_{++}\ca-\ca\Lambda\inv_{++}
-\ca\cb\Lambda\inv_{-+}) \nonumber\\
& = & \Lambda^{1-k}_{++}\partial_{-1}\ca,\,k\geq1.
\EEA
It follows that the evolution of the upper diagonal block w.r.t.\ an arbitrary
flow is determined by its evolution w.r.t.\ the first negative and the first
positive flow.

Equations \bref{cmhier} and \bref{bphier} make it possible to express
$\partial_k\cc_{-1,j}$ and $\partial_{-k}\cb_{j,-1}$ for $j,k\geq1$ as
differential polynomials in $\cc_{-1,0}$ and $\cb_{0,-1}$,
respectively. The question arises, if this is also possible for the
other partial derivatives $\partial_{-k}\cc_{-1,j}$ and
$\partial_k\cb_{j,-1}$.

Using the commutativity of the flows, we can evaluate
$\tau(m_\xi+n_\zeta,g(\bt))$ in two different ways.
We get by \bref{groupprop} and Taylor expansion the following set of
equations ($i,j\geq1$):
\BEA
\lefteqn{P_j\cb_{i-1,-1}(\bt)
-\sum_{k=1}^{j-1}\cc_{-1,k-1}(\bt)P_{j-k}\cb_{i-1,-1}(\bt)} & & \nonumber\\
& = & P_{-i}\cc_{-1,j-1}(\bt)
-\sum_{k=1}^{i-1}\cb_{k-1,-1}(\bt)P_{k-i}\cc_{-1,j-1}(\bt).
\EEA
This doesn't allow us to express the negative derivatives of
$\cc_{-1,j+1}$ by derivatives of $\cc_{-1,0}$, however, it gives us
new relations expressing for example $\partial_{-k}\cc_{-1,j}$ as a
polynomial in $\cc_{-1,i}$, $i<j$ and positive derivatives of $\cb_{0,-1}$.

Let us abbreviate
$u\DEF\cc_{-1,0}$, $v\DEF\cb_{0,-1}$.
If we assign the following degrees to the matrix entries and
derivation operators:
\BEA \label{grading}
\deg(\cb_{j,k}) & = & \deg(\cc_{j,k})=k-j,\\
\deg(\partial_m) & = & m.
\EEA
then the equations \bref{cceqcoord}, \bref{cbeqcoord},
\bref{bphier} and \bref{cmhier} respect this gradation.

The picture that develops here is the following: We have in principle
two copies of the KP-hierarchy, given by the component $u=\cc_{-1,0}$
together with the positive flows and $v=\cb_{0,-1}$ and the negative
flows, respectively. They are not completely independent, but are
coupled by the equation \bref{bcrel}.
Therefore the PKP-hierarchy for $v$ can be pulled back to yield
additional differential equations in $u$ and $v$, commuting with the
PKP-hierarchy for $u$. These equations are belonging to the $t_k$ flows
for negative $k$.

For example the $t_{-3}$-flow gives the PKP-equation for $v$:
\BEA
{3\over4}\partial_{-2}^2v
& = & -\partial_{-2}^2\partial_{-1}\ln\tau \nonumber\\
& = & \partial_{-2}\partial_{-1}(\cb_{1,-1}+\cb_{0,-2}) \nonumber\\
& = & \partial_{-1}(\partial_{-3}v-{1\over4}\partial_{-1}^3 v
+{3\over2}(\partial_{-1}v)^2),
\EEA
by equation \bref{bphier}.
For $u$ this yields:
\BEA
\partial_{-1}\partial_{-2}^2u & = & \partial_{-2}^2\partial_1 v \nonumber\\
& = & \partial_{-1}\partial_1({4\over3}\partial_{-3}v
-{1\over3}\partial_{-1}^3 v+(\partial_{-1}v)^2).
\EEA
Neglecting an integration constant we get the equation in $u$ and $v$:
\BEQ
{3\over4}\partial_{-2}^2 u=-\partial_{-1}\partial_{-3}u
+{1\over4}\partial_{-1}^4 u-{3\over2}(\partial_{-1}^2u \partial_{-1}v).
\EEQ

\section{The set of solutions} \label{solsec}
Up to now it may not be clear, why we did not start with a group like
$\GLRES(k)$ in the first place, thereby avoiding the analytic difficulties
we encounter with the infinite determinant and the definition of the
$\tau$-function.

First we will see in the next section, that the reduction to $1+1$-dimensional
equations like the KdV-equation requires the full group $\GLRES$ or at least a
dense subset in it.

Second it is hard to see, which solutions we throw away by looking at the
restricted subgroups. We will see that, in this setting unexpectedly,
we encounter the appearance of another splitting, the analog of the
well known Birkhoff factorization of $\GLRES$.

The question we are primarily interested in is, what initial conditions give
the same $\tau$-function. This amounts to the question how far the diagonal
blocks of the initial matrix are determined by the offdiagonal blocks.
The latter ones are given by the $\tau$-function. On the other hand
for given $\cc_{-1,0}$ and $\cb_{0,-1}$ all derivatives
$\partial_k\cc_{-1,j}$ and $\partial_{-k}\cb_{j,-1}$, $k\geq1$ and
$j\geq1$ are given by the equations \bref{cmhier} and \bref{bphier} as
differential polynomials in $\cc_{-1,0}$ and $\cb_{0,-1}$.
The Riccati equations \bref{cceqcoord} and \bref{cbeqcoord} then give
the whole offdiagonal blocks, and therefore all derivatives of
$\ln\tau$ are determined up to integration constants. We can
fix these by requiring
\BEQ \label{cond}
\partial_k\ln\tau(0,g_0)=0 \label{initial}
\EEQ
for all $k$ with $|k|>1$, and all splittable $g_0$.

If we restrict ourselves to the subgroup $\GLRES^{(2)}$ of $\GLRES$, for which
the
offdiagonal blocks are Hilbert-Schmidt, then
we may actually show, that for all splittable $g_0$ the condition \bref{cond}
can be enforced by acting
with the group $\Gamma_-\times\Gamma_+$: Denoting with $L_{\gamma_-}$ and
$R_{\gamma_+}$ the left multiplication with
$\gamma_-=\exp\sum_{k<0}t_k\Lambda^k\in\Gamma_-$ and right
multiplication with $\gamma_+=\exp\sum_{k>0}t_k\Lambda^k\in\Gamma_+$,
respectively, we see immediatly:
\BEA
\tau(g_+,L_{\gamma_-}g_0) & = & c(\gamma_-,g_+)\tau(g_+,g_0),\\
\tau(g_-,R_{\gamma_+}g_0) & = & c(g_-,\gamma_+)\tau(g_-,g_0),
\EEA
and therefore for $k>1$:
\BEA
\partial_k\ln\tau(0,L_{\gamma_-}g_0) & = & \partial_k\ln\tau(0,g_0)-kt_k,\\
\partial_-k\ln\tau(0,R_{\gamma_+}g_0) & = & \partial_k\ln\tau(0,g_0)-kt_{-k}.
\EEA
Choosing $t_k$ appropriately for all $k$, $|k|>1$, we can reach
\bref{initial} from any initial condition by acting with a uniquely but
formally defined element $\gamma_-\times\gamma_+$
in $\Gamma_-\times\Gamma_+$ in the prescribed way.
In the case, where the offdiagonal blocks of $g_0$ are Hilbert-Schmidt, the
$\gamma_+$ and $\gamma_-$ obtained by this procedure converge in $\Gamma_+$
and $\Gamma_-$. The same holds, as was shown in \cite{DPL1}, if one works with
certain weighted $\ell_1$ Banach structures instead of $\GLRES$.
For compact offdiagonal blocks, on the other hand, one can employ the example
of Pressley and Segal \cite[(8.3.4)]{PS} to show, that the offdiagonal blocks
of $\gamma_-\times\gamma_+$ may not even be bounded operators.

The ambiguity left is an overall multiplicative constant, which we can
set to be one, since it doesn't affect the logarithmic derivatives of $\tau$.

We look at the differential equations \bref{cbeq} and \bref{cceq} for
the offdiagonal blocks $\cb$ and $\cc$, respectively.
It is obvious, that they, and therefore the solutions, don't change, if the
last terms containing $\ca$ and $\cd$ do not.
This gives us conditions on the stabilizer of a solution.
In the following we will consider only splittable initial conditions.
We write out the conditions on the last terms in \bref{cbeq} and \bref{cceq}
as matrix equations:

Let $\oa$, $\od$ and $\ca$, $\cd$ be pairs of matrices, for which
\BEA
\oa\inv\Lambda^m_{+-}\od & = & \ca\inv\Lambda^m_{+-}\cd, \label{bstab}\\
\od\Lambda^m_{-+}\oa\inv & = & \cd\Lambda^m_{-+}\ca\inv. \label{cstab}
\EEA
We may write
$$
D=\od\inv\cd,\,A=\oa\inv\ca,
$$
and thus obtain from equation~\bref{cstab} the condition
\BEQ
D\Lambda^{-m}_{-+}=\Lambda^{-m}_{-+}A.
\EEQ
In coordinates this is
\BEA
(D\Lambda^{-m}_{-+})_{ij} & = & \Case{D_{i,j-m}}{j=0,\ldots,m-1}{0}{j\geq m},
i<0, \\
(\Lambda^{-m}_{-+}A)_{ij} & = & \Case{A_{i+m,j}}{i=-m,\ldots,-1}{0}{i<-m},
j>0.
\EEA
Therefore $A$ and $D$ are lower echelon matrices, which are determined by
each other.

The notion ``triangular'' here means that the entries on the diagonal are
all $1$, an echelon matrix is a product of an arbitrary diagonal and a
triangular matrix.

In the same way we derive from equation~\bref{bstab} the conditions
\BEQ
\tA\Lambda^m_{+-}=\Lambda^m_{+-}\tD
\EEQ
for the matrices
$$
\tD=\cd\od\inv,\,\tA=\ca\oa\inv,
$$
which, because of
\BEA
(\Lambda^m_{+-}\tD)_{ij} & = & \Case{\tD_{i-m,j}}{i=0,\ldots,m-1}{0}{i\geq m},
j<0, \\
(\tA\Lambda^m_{+-})_{ij} & = & \Case{\tA_{i,j+m}}{j=-m,\ldots,-1}{0}{j<-m},
i>0,
\EEA
amount to $\tA$ and $\tD$ being upper echelon matrices, which are
determined by each other.

We end up with the following condition:
The set of $g\in\GLRES$ yielding the same $\tau$-function as
$$
\tmatrix{\unity}{0}{\cc}{\unity}\tmatrix{\ca}{0}{0}{\cd}
\tmatrix{\unity}{\cb}{0}{\unity},
$$
and for which \bref{initial} holds,
is parametrized by invertible upper
and lower triangular Fredholm matrices $\tA$ and $A$, for which
\BEQ
\ca=\tA\ca A\inv. \label{stab}
\EEQ
We showed before, that if we look at $\GLRES^{(2)}$,
$\tGamma=\Gamma_+\times\Gamma_-$
acts freely on the matrices $g_0$, which give the same solution.
Let's denote by
$\caS$ the quotient of the flow action of $\Gamma$ and the above described
action of $\tGamma$ on $\GLRES^{(2)}$. A solution then determines
an element of $\caS$ up to a transformation of the upper triangular block by
\bref{stab}. I.e.\ the fibre in $\caS$ over a solution, which has an initial
condition with upper diagonal block $\ca$ is given by the
stabilizer of $\ca$ w.r.t.\ the action \bref{stab} of invertible upper and
lower echelon Fredholm operators.

Notice that the stabilizer of $\ca$ may
vary in a very complicated way over $\GLRES$.
We want to investigate the structure of the stabilizer in a special
case a little
further. We employ the Gauss algorithm in order to split the matrix
$\ca$ into an upper triangular, a diagonal and a lower triangular matrix.

The Gauss algorithm allows us to write down matrices $\ta_L$, $a_D$ and
$a_U$, s.t.\
$$
\ca \ta_L= a_U a_D,
$$
for $\ca$ in a dense subset of $B(H_+)$, to which we restrict ourselves.
Even for infinite matrices it is no problem to multiply a lower
triangular matrix like $a_L=\ta_L\inv$ from the right with an arbitrary
matrix, as all matrix elements of the product consist of sums
of finitely many products of matrix entries.
Therefore we have a kind of Birkhoff factorization
\BEQ \label{birk}
\ca = a_U a_D a_L,
\EEQ
where the factors are uniquely determined matrices, but not necessarily
bounded operators on $H_+$.

Writing down \bref{stab} we get
$$
a_U\inv \tA a_U a_D a_L A\inv a_L\inv=a_D.
$$
The only freedom we have, is to choose the diagonal part of
$a_U\inv \tA a_U$, because of the
uniqueness of the splitting \bref{birk}.
However, as $\tA$ has to be a bounded operator, we get additional
highly nontrivial conditions on this diagonal matrix.
To be precise, if
$$
a_U \inv \tA a_U=U\cdot D,
$$
$U$ being upper triangular and $D$ being diagonal, then we have to
choose $D$ in such a way, that $a_U U D a_U\inv$ is a bounded
operator.

If $\ca$ contains a permutation matrix in its Birkhoff decomposition,
then the stabilizer get's bigger, including upper triangular matrices,
which are transformed by $\ca$ to lower triangular matrices.
It is an open problem, if there is any ``solution manifold'' as in the
Grassmannian case \cite{DPL1}.

\section{A comment on the connection with the Hirota-Satsuma equation}
\label{HSsec}
We want to link up the negative PKP-hierarchy with work done by Bogoyavlenskii
and others \cite{bog90,szmig93,hirsat76,AKNS74}.

To this end we consider the usual reduction from the KP to the KdV-hierarchy
\cite{SW85}:\\
Let $\GLKdV$ be the subgroup of $\GLRES$ of all matrices, which commute with
the square of the double shift $\Lambda^2$. Taken as initial conditions of the
flows~\bref{action}, these are precisely the ones on which all even
numbered flows act trivially.

This subgroup can easily be identified with the loop group
$L\LieGL(2,\Bcc)$, where the identification in terms of matrix units is
\BEQ
e_{ij}\longrightarrow \lambda^{\left[{i\over 2}\right]
-\left[{j\over 2}\right]}e_{i\,\mod\, 2,j\,\mod\, 2},
\EEQ
$[x]$ denoting the greatest integer less than $x$.
This way the double shift $\Lambda$ is identified with the matrix
\BEQ
p=\tmatrix{0}{1}{\lambda}{0}.
\EEQ
The identification, however, does not respect the splitting~\bref{splitting}
as it identifies offdiagonal with diagonal blocks.
As we are not really interested in the behaviour of the diagonal
blocks we include them into the positive part.

This yields a well known
formulation \cite{DPKDV} of the potential KdV-equation (PKdV) in terms of a
loop group splitting, the PKdV thus being the standard loop group reduction
of the potential KP-hierarchy.

Again, we let the flows act by conjugation.
The subgroups of the splitting are now
\BEA
G_+ & = & \{g(\lambda)\in L\LieGL(2,\Bcc) | g(\lambda)\,
\mbox{\rm is analytic inside the unit circle}\}, \\
G_- & = & \{g(\lambda)\in L\LieGL(2,\Bcc) | g(\lambda)\,
\mbox{\rm is analytic outside the unit circle}, \nonumber\\
& & \kern2mm g(0)=\unity\},
\EEA
The product $G_-G_+$ is a dense open subset in $L\LieGL(2,\Bcc)$ \cite{PS}.
Since the independent function
$u=\cc_{-1,0}$ which satisfies the extended PKP-hierarchy
is mapped to the upper right corner of the $\lambda^{-1}$ coefficient
of $g_-$, the latter is also the function which satisfies the
reduced negative KP-hierarchy.

The first equation plays the same role for the KdV-equation as the sine-Gordon
equation does for the modified KdV-equation.

The flow matrices are mapped to the matrix
\BEA
\Lambda & \longrightarrow & p_1=\tmatrix{0}{1}{\lambda}{0}, \\
\Lambda\inv & \longrightarrow & p_{-1}=\lambda^{-1}p_1.
\EEA
Therefore the flows read
\BEQ \label{flowNKdV}
g(x,t)=\exp(xp_1+tp_{-1})g_0\exp(-xp_1-tp_{-1})=g_-(x,t)g_+(x,t)\inv.
\EEQ
To reproduce a zero curvature condition for the negative flows we
adopt the double group formulation of negative flows~\cite{Wu,Haak}
This amounts to the following: If we look at \bref{flowNKdV} we may
write the splitting in a slightly different way, $g_0=g_{0-}g_{0+}\inv$,
\BEA
g(x,t) & = & (\exp(xp_1+tp_{-1}),\exp(xp_1+tp_{-1}))(g_{0-},g_{0+})
\nonumber\\
& = & (g_-(x,t),g_+(x,t))(v(x,t),v(x,t)),
\EEA
where we have identified the set of splittable matrices in
$L\LieGL(2,\Bcc)$ to the subset $G_-\times G_+$ of the product group
$L\LieGL(2,\Bcc)\times L\LieGL(2,\Bcc)$ by
\BEQ
g_-g_+\inv \longrightarrow (g_-,g_+).
\EEQ
The additional factor $(v,v)$ occurs due to the fact, that the flows
do not stay in the subgroup $G_-\times G_+$.
This amounts to a splitting of the double group $\calG=L\LieGL(2,\Bcc)
\times L\LieGL(2,\Bcc)$
into the subgroups $\calG_-=G_-\times G_+$ and
$\calG_+=\diag(\calG)=\{(x,y)\in\calG|x=y\}$.
The flows then act by left multiplication with
$(\exp(xp_1+tp_{-1}),\exp(xp_1+tp_{-1}))\in\calG_+$ and reproduce the
well known formulation of an integrable system with only positive flows.

We explicitly write down $v(x,t)$:
\BEA \label{veq}
v(x,t) & = & g_-(x,t)\inv \exp(xp_1+tp_{-1})g_{0-} \nonumber\\
& = & g_+(x,t)\inv \exp(xp_1+tp_{-1})g_{0+}.
\EEA
It follows that the matrices
\BEA
U & = & \partial_xv\,v\inv, \\
V & = & \partial_tv\,v\inv,
\EEA
satisfy the zero curvature condition
\BEQ \label{ZCC}
\partial_tU-\partial_xV+[U,V]=0.
\EEQ
We further reduce to the loop group $LSU(2)$ and write
\BEQ
g_-=\exp(\lambda\inv\tmatrix{a}{b}{c}{-a}+\mbox{\rm lower order terms}).
\EEQ
We therefore get
\BEQ
U(x,t)=(\Ad(g_-\inv)p_1)_+=\tmatrix{-b}{1}{\lambda+2a}{b}
\EEQ
and by
\BEQ
V(x,t)_-=(\Ad(g_+\inv)p_{-1})_-=\lambda\inv V_{-1},
\EEQ
with $V_{-1}$ independent of $\lambda$, also
\BEA
V(x,t) & = & \tmatrix{0}{0}{1}{0}+(-g_-\inv\partial_tg_-+g_-\inv
p_{-1}g_-)_-
\nonumber\\
& = & \lambda\inv U(x,t)-\tmatrix{a_t}{b_t}{c_t}{-a_t}\lambda\inv.
\EEA
Here the subscripts $+$ and $-$ indicate projection to the Lie algebras of
$\GLP$ and $\GLM$, respectivly.

In addition we have by evaluating the upper right corner
of the $\lambda\inv$-coefficient of
\BEQ \label{gmineq}
g_-\inv\partial_xg_--\Ad(g_-\inv)p_1=g_+\inv\partial_xg_+-\Ad(g_+\inv)p_1
\EEQ
$a=-{1\over2}(b_x+b^2)$. This is the Riccati equation \bref{cceqcoord} for the
reduced case, where $\cc_{-1,1}+\cc_{-2,0}=\partial_2\ln(\tau)=0$.

Evaluating \bref{ZCC} yields the first negative equation in the
KdV-hierarchy for the matrix element $b$:
\BEQ
b_{xx}-{1\over4}b_{xxxt}-2b_xb_{tx}-b_tb_{xx}=0,
\EEQ
or for $u(x,t)=-4(b(x,t)-t)$:
\BEQ
u_{xxxt}=u_tu_{xx}+2u_xu_{xt},
\EEQ
which is Bogoyavlenskii's version of the Hirota-Satsuma equation
\cite{hirsat76} for long
waves in a medium with nonlinear dispersion. Bogoyavlenskii \cite{bog90}
investigated it as an integrable equation with overturning (breaking) solitons.
In the setting of this section it was derived as the first negative flow
of the potential KdV-equation by Szmigielski \cite{szmig93}.

In the case of the Grassmannian formulation of the potential KP equation, one
recovers the elements of the potential KdV hierarchy as reductions of
equations of the PKP-hierarchy, i.e.~simply by setting derivatives w.r.t.~even
numbered variables to zero.

This works even though the splitting in the reduced and the unreduced case
is quite different, i.e.~in the reduced case we split along the diagonal, in
the unreduced case we leave whole block matrices on the diagonal.

In the reduction the $\tau$-function gets lost and is replaced by the zero
curvature condition. This yields the fact, that due to the loss of the
quations~\bref{cmhier} and \bref{bphier} in the reduced case, we can no more
express the offdiagonal
blocks merely by one corner entry $\cc_{-1,0}$ or $\cb_{0,-1}$, respectively.

Instead we end up in the reduction to $LSU(n)$ with $n-1$ seemingly
independent functions, which describe the offdiagonal blocks. The Riccati
equations \bref{cceqcoord} and \bref{cbeqcoord} are still applicable and occur
in the form of equation~\bref{gmineq}.

We plan to investigate the connection between the extended PKdV-hie\-rar\-chy
and the extended PKP-hierarchy in a separate publication.


\begin{thebibliography}{DNS99}
\frenchspacing
\bibitem{SW85} G. Segal, G. Wilson, ``Loop Groups and Equations of KdV type'',
Publ. IHES 61, 1985.

\bibitem{DNS89} J. Dorfmeister, E. Neher, J. Szmigielski, ``Automorphisms of
Banach Manifolds Associated with the KP-equation'', Quart. J. Math. Oxford (2),
40 (1989).

\bibitem{DS} V. G. Drinfeld, V. V. Sokolov, ``Equations of Korteweg-de Vries
type and simple Lie algebras'', Soviet Math. Dokl. 23 (1981).

\bibitem{AvM} M. Adler, P. van Moerbeke, ``Completely integrable
systems, Euclidean Lie algebras and curves'' and ``Linearization of
Hamiltonian systems, Jacobi varieties and representation theory'',
Adv. Math. 38 (1980).

\bibitem{simon} B. Simon, ``Trace Ideals and Their Applications'', London
Math. Soc. Lecture Notes 35, Cambridge University Press, 1979.

\bibitem{bog90} O. I. Bogoyavlenskii, ``Overturning Solitons in New
Two-Di\-men\-sional Integrable Equations'', Math. USSR Izvestiya 34
(1990).

\bibitem{szmig93} J. Szmigielski, ``On Soliton Content of Self Dual
Yang-Mills Equations'', Preprint hep-th 9311119, 1993.

\bibitem{hirsat76} R. Hirota, J. Satsuma, ``N-soliton of Model
Equations for Shallow Water Waves'', J. Phys. Soc. Japan 40 (1976).

\bibitem{AKNS74} M. J. Ablowitz, D. J. Kaup, A. C. Newell, H. Segur, Stud.
Appl. Math. 53, 249 (1974).

\bibitem{DPKDV} J. Dorfmeister, ``Banach Manifolds of Solutions to
Nonlinear
Partial Differential Equations, and Relations with Finite Dimensional
Manifolds''.

\bibitem{DPL1} J. Dorfmeister, ``Weighted $\ell_1$-Grassmannians and
Banach Manifolds of Solutions to the KP-equation and the KdV-equation''.

\bibitem{Haak} G. Haak, ``Analytische Untersuchungen an klassischen und
quantenintegrablen Systemen'', PhD-thesis, Freie Universit\"at Berlin,
FB Physik, 1993.

\bibitem{Wu} H.-Y. Wu, ``Nonlinear partial differential equations via
vector fields on homogenous Banach manifolds'', Ann. Global Anal. Geom.
10 (1992).

\bibitem{PS} A. Pressley, G. B. Segal, "`Loop Groups"',
Clarendon Press, Oxford, 1986.
\end{thebibliography}
\end{document}